\documentclass[preprint,showpacs,showkeys,preprintnumbers,amsmath,amssymb]{revtex4-1}
\usepackage{graphicx}
\usepackage{dcolumn}
\usepackage{bm}

\begin{document}
\input{epsf}


\title{Weakly bound states of neutrons in gravitational fields\\}
\author{Avas V. Khugaev\footnote{Permanent address: Institute of Nuclear Physics, 100214 Tashkent, Uzbekistan}}%
\email{avaskhugaev@yahoo.com}
\affiliation{Bogoliubov Laboratory of Theoretical Physics, \\
Joint Institute of Nuclear Research, 141980 Dubna, Russia}
\author{Renat A. Sultanov}
\email{rasultanov@stcloudstate.edu}
\affiliation{Department of Information Systems and BCRL,
St. Cloud State University,
367B Centennial Hall, 720-4th Avenue South, St. Cloud, MN 56301-4498, USA}
\author{Dennis Guster}
\email{dcguster@stcloudstate.edu}
\affiliation{Department of Information Systems and BCRL,
St. Cloud State University,
367C Centennial Hall, 720-4th Avenue South, St. Cloud, MN 56301-4498, USA
}%

\date{\today}

\begin{abstract}
In this paper a quantum-mechanical behaviour of neutrons in gravitational fields is considered.
A first estimation is made using the semiclassical approximation, neglecting General Relativity, magnetic
and rotation effects, for neutrons in weakly bound states in the weak gravitational field of the Earth.
This result was generalized for a case, in which the Randall - Sundrum  correction to Newton's
gravitational law on the small scales was applied. Application of the results to Neutron Star physics is considered and
further possible perspectives are discussed.
\end{abstract}

\pacs{03.65.Ta, 98.80.Cq}

\maketitle

\section{\label{Intro}Introduction
}

The idea concerning the potential of neutron storage by using of UCN (ultra cold neutrons), probably was
first suggested in 1959 by Ya.B. Zeldovich \cite{Zeldovich}. In his work a simple estimate concerning the
possibility of UCN conservation in the container was devised. It was shown that a UCN
(with wave length larger than 500 $\dot{A}$ and effective temperature less than $10^{-3}$ K)
should be totally reflected from the reservoir walls (which were made from either carbon or beryllium material)
when the neutrons are at velocities around $5 m/s$ and less.

More rigorous results, obtained later, completely confirmed these estimations. In general, these
results can be obtained in the framework of pure quantum - mechanical or optical approaches. In the quantum
mechanical approach the potential of the reflecting surface is constructed on the basis of the average
procedure of the pointlike Fermi quasipotential, given in the form:
\begin{eqnarray}
u(r)=\frac{2\pi\hbar}{m} \cdot b\delta (\vec r -\vec r_i)\nonumber
\end{eqnarray}
where $b$ - is a neutron wave length of coherent scattering on the nuclei of the considered surface.
In this approach, by strictly following the framework of quantum mechanics it was shown that the depth of neutron
penetration  in the surface material are significantly less, than the neutron wave length
\cite{Sapiro1,FrankA1,FrankA2} and the neutron scattering on the surface
can be considered as elastic. An alternative approach \cite{FrankI1}, was based on the
optical analogy of light scattering on metals, such that all inelastic processes in the
reflection can be effectively described by the imaginary part of the refraction coefficient.
In this case the work of \cite{Zeldovich} serves a necessary theoretical premise for the starting point
of all further experimental works based on the UCN technique.
One of the first experimental results was obtained in 1968 by the JINR (Joint Institute of Nuclear Research,
Dubna)\cite{Luschikov}. The importance of this type of  work is extremely high, because it opens
a new and unique possibility to carry out highly precise experiments, such as accurate
measurement of neutron lifetime or its electrical dipole moment \cite{Sapiro2} and etc.

The first proposal using experimental measurements with UCN in the gravitation field
of the Earth was done by \cite{LuschikovFrank}.
Therefore, there appears to be sound knowledge base from which to proceed given the achievements to date
in both the theoretical and experimental fields. Certainly, the many recent works devoted to this
research subject indicate that the solution of the UCN storage problem can give us an unique capability
to carry out a very precise experiments in the field of neutron physics and neutron
interferometry in gravitational fields \cite{Colella,Rauch}. In this paper, we will begin by reviewing
some pertinent recent experimental results and measurements of UCN weakly bound quantum states
in  the weak gravitational field of the Earth \cite{Nesvizh1,Nesvizh2,Nesvizh3,Nesvizh4,Nesvizh5}.
After that, we will present a simple sketch of some of the other possible applications of these results
to astrophysics and gravity. A more detailed description of some of other applications can be found
in \cite {Nesvizh6}.

\section{ Energy of weakly bound states of neutrons in the Earth's gravitational field.}

Theoretical consideration of a neutron's energy spectrum and its wave functions in the Earth's
gravitational field, as cited in the \cite{Nesvizh5}, can be found in a number of
works, including, for example \cite{Goldman, Sakurai, Davies}. In this section, by using pure
methodic reasoning, we present a simple and transparent derivation of the main relationships of
these results. A description of weakly bound states in the gravitational field of the Earth follows in
the Schr\"odinger Equation (SE) solution using a simple spherically symmetric case with
gravitational potential energy, defined as:
\begin{displaymath}
\quad\delta V(r)=
\left\{\begin{array}{ll}
-\gamma Mm\biggl(\frac{1}{r}-\frac{1}{R_{0}}\biggr), & r > R_{0}        \\
                         & \qquad\qquad\qquad\qquad                     (1)\\
\infty, & 0\leq r\leq R_{0}
\end{array}\right.
\end{displaymath}
where $M$ and $m$ are the Earth and neutron masses at rest respectively and $r$ - is a distance
from the Earth in cm
up to the point where the neutron mass $m$ is placed. In this case SE can
be written in the following way:
\setcounter{equation}{1}
\begin{eqnarray}
\frac{1}{r^2}\frac{d}{d r}
\Biggl(r^{2}\frac{d\psi}{d r}\Biggr)+
\frac{2m}{\hbar^{2}}\biggl[E-\delta V(r)\biggr]\psi
=0
\end{eqnarray}
where $\psi$ is a neutron wave function and $E$ is its corresponding energy.
Substituting, in the normal way, $\psi=\frac{f(r)}{r}$ into the equation $(2)$ we have:
\begin{eqnarray}
\frac{d^{2}f}{dr^2}+
\frac{2m}{\hbar^{2}}\biggl[E-\delta V(r)\biggr]f
=0.
\end{eqnarray}
Using a simple approximation for the gravitational potential energy in the vicinity
of the Earth surface, we can write:
\begin{eqnarray}
\delta V(r) \to \delta V(x)=\gamma \frac{Mm}{R_{0}}\Biggl[\frac{x}{R_{0}}+o\Biggl(\frac{x}{R_{0}}\Biggr)^2\Biggr]
\end{eqnarray}
where $r=R_0+x$, $\frac{x}{R_0}\ll 1$ and $R_0$ is an average radius of the Earth. In this
approximation SE can be written as:
\begin{eqnarray}
\frac{d^{2}f}{dx^2}+(\alpha - \beta x)f=0,
\end{eqnarray}
where we introduced $\alpha$ and $\beta$ as parameters, defined by:
\begin{eqnarray}
\alpha=\frac{2m}{\hbar^2}E; \quad \beta=2\Biggl(\frac{m}{\hbar} \Biggr)^2g; \quad g=\gamma\frac{M}{R^{2}_{0}}
\end{eqnarray}
By substituting a new variable $z=\alpha - \beta x$, we can finally write SE in the convenient form
for the analytic solution:
\begin{eqnarray}
\frac{d^{2}f}{dz^2} - \frac{z}{(i\beta)^2}f=0
\end{eqnarray}
For this ordinary differential equation of the second order, as suggested by the theory of Bessel
functions \cite{Watson}, there exists a regular solution which is presented in the following form:
\begin{eqnarray}
f(z)=c_{N}\sqrt{z}\Biggl[J_{\frac{1}{3}}\Biggl(\frac{2z^{\frac{3}{2}}}{3\beta}\Biggr)+
J_{-\frac{1}{3}}\Biggl(\frac{2z^{\frac{3}{2}}}{3\beta}\Biggr)\Biggr]
\end{eqnarray}
where $c_N$ is a normalization coefficient of the wave function. From this last expression
we can extract the energy spectrum of the neutrons weakly bound states by using a boundary condition
at the point $x=0$, which simply leads to the equation:
\begin{eqnarray}
J_{\frac{1}{3}}\Biggl(\frac{2\alpha^{\frac{3}{2}}}{3\beta}\Biggr)+
J_{-\frac{1}{3}}\Biggl(\frac{2\alpha^{\frac{3}{2}}}{3\beta}\Biggr)=0
\end{eqnarray}
If you assume that $\zeta_n\to \zeta =\frac{2\alpha^{\frac{3}{2}}}{3\beta}$ is such a number,
then it satisfies the equation:
\begin{eqnarray}
J_{\frac{1}{3}}\biggl(\zeta_n\biggr)+J_{-\frac{1}{3}}\biggl(\zeta_n\biggr)=0
\end{eqnarray}
which leads us to:
\begin{eqnarray}
E_n=\Biggl(\frac{9m}{8} (\hbar g \zeta_n )^2\Biggr)^{\frac{1}{3}}=c_{0}(\zeta_n)^{\frac{2}{3}}
\end{eqnarray}
where we introduced $c_{0}= \Biggl(\frac{9m}{8} (\hbar g )^2\Biggr)^{\frac{1}{3}}$. By now using
approximate values for the above defined constants, such as: $g\approx 9.80655$ m${s^{-2}}$;
$R_0\approx 6.371\cdot 10^6$ m; $mc^2\approx 939.565330$ MeV and
$\hbar c\approx 197.327053$ MeV$\cdot$ fm, we can calculate the first bound states for
the $E_n$ values and compare them with other results, including experimental measurements.
These results are presented in Table 1, where the $c_0$ value is set to $c_0\approx 0.757325\cdot 10^{-12}$ eV.
\begin{table}
\caption{\label{tab:table1}Comparison of our results with theoretical and experimental
results, obtained in \cite{Nesvizh3}. Energy values are given in peV units and the $\zeta_n$ value
is dimensionless and $E_n$-th is calculated using expression $(17)$ }
\begin{ruledtabular}
\begin{tabular}{cclcl}
n &$\zeta_n$&$E_n$-th&$E_n$,\cite{Nesvizh3} -exp.&
$E_n$, our \\ \hline
1 & 2.3834   & 1.3767  & 1.4 & 1.4054 \\
2 & 5.5105   & 2.42191 & 2.5 & 2.4573 \\
3 & 8.6474   & 3.27356 & 3.3 & 3.3180 \\
4 & 11.7868  & 4.0255  & 4.1 & 4.0794 \\
5 & 14.9272  & 4.71261 & $-$ & 4.7751 \\
\end{tabular}
\end{ruledtabular}
\end{table}
\space{10mm}
In Fig.1 an approximate distribution of the $\zeta_n$ roots is presented.
\begin{figure}
\includegraphics{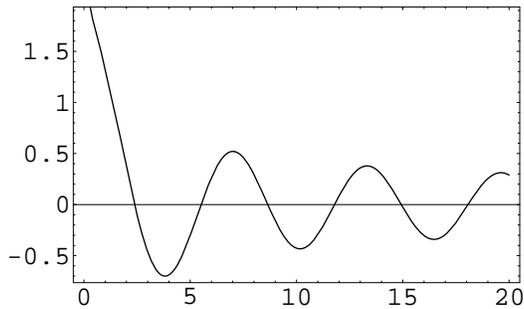}
\caption{\label{fig:epsart}$\zeta _n$ roots distribution for the
$J_{\frac{1}{3}}(\zeta_n)+J_{-\frac{1}{3}}(\zeta_n)=0$ equation.}
\end{figure}

\section{Asymptotic properties}

Given the solution obtained above it appears that it would be interesting to analyze its asymptotic
properties. To do this we are able to use the properities of Bessel functions \cite{Watson}:
\begin{eqnarray}
J_\nu(y)=\biggl(\frac{2}{\pi y}\biggr)^{\frac{1}{2}}\cos \biggl[y-\frac{\pi}{2}(\nu+\frac{1}{2})\biggr]
\end{eqnarray}
by introducing $y\to \frac{2z^{\frac{3}{2}}}{3\beta}$ we can obtain:
\begin{eqnarray}
f(y)\to 3^{\frac{5}{6}}\sqrt{\pi}\beta^{\frac{1}{3}}y^{-\frac{1}{6}}\sin{\biggl[y+\frac{\pi}{4}\biggr]}
\end{eqnarray}
from which we arrive at:
\begin{eqnarray}
f(z)=c_N\cdot 2^{\frac{1}{3}}\biggl(\frac{\pi\beta}{2}\biggr)^{\frac{1}{2}}\cdot z^{-\frac{1}{4}}
\sin{\biggl[\frac{2\alpha^{\frac{3}{2}}}{3\beta}+\frac{\pi}{4}\biggr]}
\end{eqnarray}
For the zero roots of this function we can immediately obtain from
the last expression that:
\begin{eqnarray}
\frac{2\alpha^{\frac{3}{2}}}{3\beta}+\frac{\pi}{4}=n\pi
\end{eqnarray}
and finally, using our notations for variable $z$ we are led to:
\begin{eqnarray}
E_n=mgx+\biggl( \frac{9m}{8}(\pi\hbar g)^2 \biggr)^{\frac{1}{3}}\biggl(n-\frac{1}{4}\biggr)^{\frac{2}{3}}
\end{eqnarray}
Naturally, one might find the last expression to be quite interesting. This is true because the 1-st term
of the expression can be precisely viewed as the classical part of an energy particle in a homogeneous
gravitational field while the second term is a quantum mechanical contribution. In the case where
$\hbar\to 0$ we have arrived at a pure classical result and in the case where $\hbar \neq 0$, but $x=0$
we have exactly obtained results matching the energy spectrum from \cite{Nesvizh3}:
\begin{eqnarray}
E_n=\biggl( \frac{9m}{8}(\pi\hbar g)^2 \biggr)^{\frac{1}{3}}\biggl(n-\frac{1}{4}\biggr)^{\frac{2}{3}}
\end{eqnarray}

\section{The search for high dimension contributions.}

As it was pointed out in \cite{Nesvizh3}, the possibility of making observations
concerning the UCN bound states in the Earth's weak gravitational field can provide
new insight regarding the verification of known interactions involving small
distances.

An interesting view on the gravitation interaction, theoretically can lead us
to the idea, elegantly declared in \cite {Naresh}, regarding its universality and
multi-dimension nature. In particular, it follows from  \cite{RS} that a high
dimensional correction using the framework of the Randall - Sundrum (RS) theory can be applied
to Newton's gravitational law at small distances. To obtain a better understnding of this premise
it is useful to evaluate such theoretical predictions from the experimental
point of view. An interesting proposal related to the search of the RS correction to
Newton's gravitational law on small scales was recently suggested by \cite{Sami}.

In this section our goal is to theoretically estimate the possible contribution the
RS correction might have on Newton's gravitational law at small distances on the
UCN weakly bound states in the Earth's gravitational field and  to calculate an
upper limit of the $l_{RS}$ (Randall - Sundrum) parameter:
\begin{widetext}
\begin{displaymath}
\qquad\qquad\qquad\qquad\qquad\quad V(r)=
\left\{\begin{array}{ll}
-\gamma Mm
\biggl[\frac{1}{r}\biggl( 1+\frac{l^2_{RS}}{r^2}\biggr) -
\frac{1}{R_0}\biggl( 1+\frac{l^2_{RS}}{R_{0}^2}\biggr) \biggr], & r > R_{0}        \\
                         & \qquad\qquad\qquad \qquad\qquad\qquad\qquad\qquad  (18)\\
\infty, & 0\leq r\leq R_{0}
\end{array}\right.
\end{displaymath}
\end{widetext}
Using an approach simular to the one applied in the previous section we find that in a spherically
symmetric case that the equation $(3)$ can be rewritten by using the
$\frac{x}{R_0}\ll 1 $ approximation as:
\setcounter{equation}{18}
\begin{eqnarray}
\frac{d^{2}f}{dx^2}+
\frac{2m}{\hbar^{2}}\Biggl[E - mg\Biggl(1+3\cdot \frac{l^2_{RS}}{R_{0}^2}\Biggr)x\Biggr]f
=0
\end{eqnarray}
Introducing in a similar way as before we inject a new additional constant:
$\Lambda_{RS}=\frac{l_{RS}}{R_0}$, therefore the last equation $(19)$ can be rewritten as:
\begin{eqnarray}
\frac{d^{2}f}{dx^2}+(\tilde\alpha -\tilde\beta x)f=0
\end{eqnarray}
with redefinition of the $\alpha$ and $\beta$ values from equation $(5)$ it follows:
\begin{eqnarray}
\tilde\alpha\equiv \alpha =\frac{2mE}{\hbar^2}; \quad
\tilde\beta=2\Biggl(\frac{m}{\hbar}\Biggr)^{2}g\biggl(1+3\Lambda^{2}_{RS}\biggr)
\end{eqnarray}
As it was shown before the solution of the equation $(20)$ can be written
in the form:
\begin{eqnarray}
f(\tilde z)=\tilde c_{N}\sqrt{\tilde z}\Biggl[J_{\frac{1}{3}}\Biggl(\frac{2\tilde z^{\frac{3}{2}}}{3\tilde\beta}\Biggr)+
J_{-\frac{1}{3}}\Biggl(\frac{2\tilde z^{\frac{3}{2}}}{3\tilde\beta}\Biggr)\Biggr]
\end{eqnarray}
where $\tilde z = \tilde\alpha - \tilde\beta x $. Using a  simple connection between
$\beta$, and  $\tilde\beta$ values we arrive at:
\begin{eqnarray}
\tilde\beta=\beta\biggl(1+3\Lambda^{2}_{RS}\biggr)
\end{eqnarray}
further we can decribe the contribution of the $\Lambda_{RS}$ term regarding of
weakly bound states of the UCN in the Earth's gravitational field using the boundary
condition at $x=0$:
\begin{eqnarray}
J_{\frac{1}{3}}\Biggl(\frac{2\tilde\zeta^{\frac{3}{2}}}{3\tilde\beta}\Biggr)+
J_{-\frac{1}{3}}\Biggl(\frac{2\tilde\zeta^{\frac{3}{2}}}{3\tilde\beta}\Biggr)=0
\end{eqnarray}
From the previous equation by substituting a new variable
$\tilde\zeta = \Biggl(\frac{2\tilde\alpha^{\frac{3}{2}}}{3\tilde\beta}\Biggr)$
the final results can be presented as:
\begin{widetext}
\begin{eqnarray}
\tilde E_n=\Biggl(\frac{9m}{8}(g\hbar\tilde\zeta_n)^{2}(1+3\Lambda^{2}_{RS})^2\Biggr)^{\frac{1}{3}}
=c_{0}\biggl(1+
3\Lambda^{2}_{RS}\biggr)^{\frac{2}{3}}\tilde\zeta_{n}^{\frac{2}{3}}
\end{eqnarray}
\end{widetext}
where $\tilde\zeta_n$ are a root of the equation $(24)$. From this expression it follows
that $\Lambda_{RS}$ is equal to:
\begin{eqnarray}
\Lambda_{RS}=\frac{1}{\sqrt{3\tilde\zeta_n}}\biggl(\biggl(\frac{\delta\tilde E_{n}}
{c_{0}}\biggr)^{\frac{3}{2}}-1)\biggr)^{\frac{1}{2}}
\end{eqnarray}
from which we can estimate, that
\begin{eqnarray}
l_{RS}< 3.9\cdot 10^3 m
\end{eqnarray}


\section{Quantum Mechanics Applied to the NS surface.}

Currently, one actual problem of great import in the area of astrophysics is the detection the
Black Holes (BH) which could serve as an experimental proof of Einstein's gravitational theory.
Another very important problem is the development of different precise methods for the
measurement of the physical properties of  distant astrophysical objects, such as:
quasars, pulsars, gamma ray bursts, neutron stars (NS) and etc. In all these cases
it would be very useful to provide such experimental investigations by different
and completely independent physical methods. These experiments if successful would immediately
raise the validity of the experimental data and provide clarification concerning
the physical picture of the objects under investigation. Specifically, the greatest impact would
occur in BH and NS physics. Within NS physics there may be a very interesting
connection regarding the interplay of general relativity (GR) and the quantum mechanical
effects in the NS interior. The existence of the strange stars (SS) \cite{Weber}
underline that NS is a research area that is a very interesting and important
for our further understanding of the behaviour of matter at extreme conditions.
In this section we begin with a simple theoretical estimation because of the
\cite{Nesvizh1,Nesvizh2,Nesvizh3,Nesvizh4,Nesvizh5} approach, for the quantum
effects on the NS surface.  Introducing new notations: $M_{NS}$ - NS mass,
$M_{E}$ - Earth mass, we can derive using the framework of the nonrelativistic
approach of gravitational acceleration on the NS surface which is equal to:
\begin{eqnarray}
g_{NS}=\biggl(\frac{M_{NS}}{M_{E}}\biggr)\biggl( \frac{R_{E}}{R_{NS}}\biggr)^2\cdot g=\eta g
\end{eqnarray}
where  $R_{E}$ and $R_{NS}$ are the Earth's and  the NS radii respectively and $\eta$ is
defined as $\eta=\biggl(\frac{M_{NS}}{M_{E}}\biggr)\biggl( \frac{R_{E}}{R_{NS}}\biggr)^2$.
If the density of neutron matter on the NS surface is $\rho=mn_\rho$, then we can simply
estimate an average size of the cell, which would contain one neutron as:
\begin{eqnarray}
a=\biggl(\frac{m}{\rho}\biggr) ^{\frac{1}{3}}
\geq \lambda_{c}=\frac{\hbar}{m_{\pi^{0}}c}
\end{eqnarray}
where $m_{\pi^{0}}$ - is a $\pi^{0}$ meson mass, $\rho$ - is the neutron matter density on the
NS surface, $n_{\rho}$ - its concentration, $\lambda_{c}$ is the Compton wavelength and
$a$ - the size of the cell. In cases in which the cell size is larger than the
$\pi^{0}$ meson Compton wavelength, in that it is approximately equivalent to the short
range of nuclear forces, we can conclude that neutrons on the NS surface can be considered
in the first step approximation as a free gas in the external gravitational field of the NS.
This is because the average distance between them is larger than the radius of the nuclear interaction.

In our consideration, we suggest, that due to the Pauli exclusion principle there should exist, close
to the  NS surface, an analogy of the Fermi like surface, which can play the role of a mirror
as is the case of the experiments conducted with the UCN in the Earth's gravitational
field \cite{Nesvizh1,Nesvizh2,Nesvizh3,Nesvizh4,Nesvizh5}.

In this case, for rough estimation, we can simply follow the expressions derived before for
the Earth's gravitational field with small modification to obtain the final result for the NS.
The final expression, after such a modification of expressions $(11)$, would appear as follows:
\begin{eqnarray}
E^{(NS)}_n=\Biggl(\frac{9m}{8} (\alpha_n g \hbar)^2\Biggr)^{\frac{1}{3}}\cdot\eta ^{\frac{2}{3}}\to E^{(Earth)}_{n}\eta^{\frac{2}{3}}
\end{eqnarray}
Finally, a simple example to illustrate the calculations described earlier related to the NS with given parameters is provided,
the chosen parameters follow: $M_{NS}=2\cdot M_{Sun}\approx 4\cdot 10^{30}$ kg, $R_{NS}\approx 10^4$ m and
$M_{E}\approx 5.98\cdot 10^{24}$ kg, $R_{E}\approx 6.37\cdot 10^6$ m. The results of such calculations are
presented in Table 2. For these numerical data $\eta\approx 2.7\cdot 10^{-11}$.
\begin{table}
\caption{\label{tab:table2}Comparison of the energy of neutron's bound states (in {\it eV} units)
on the NS and Earth's surface}
\begin{ruledtabular}
\begin{tabular}{ccc}
n     & $E_n$ on NS         & $E_n$, on Earth (th.) \\ \hline
1     & $5.68\cdot 10^{-5}$ & $1.40\cdot 10^{-12}$  \\
2     & $9.92\cdot 10^{-5}$ & $2.46\cdot 10^{-12}$  \\
3     & $1.34\cdot 10^{-4}$ & $3.32\cdot 10^{-12}$  \\
4     & $1.65\cdot 10^{-4}$ & $4.08\cdot 10^{-12}$  \\
5     & $1.93\cdot 10^{-4}$ & $4.77\cdot 10^{-12}$  \\
\end{tabular}
\end{ruledtabular}
\end{table}
From the data in Table 2 we can conclude, that perhaps it is possible to
discuss the possibility of NS spectroscopy, where the transition wave lengths will be around
$\lambda_{tr}\sim \frac{2\pi \hbar c}{\Delta E_{tr}}\sim (0.124-1.24)$ cm, and
the according transition energy $\Delta E_{tr}\sim (10^{-5}-10^{-4})$ eV.
Details of this spectroscopic information are directly connected to the NS
physical parameters, such as the $M_{NS}$ and $R_{NS}$ values.
The probability of their spectroscopic transitions $n\to k$ can be described as an overlap of
the corresponding wave functions as described by the simple relation:
\begin{eqnarray}
\omega_{n\to k}=\biggl(\int\psi_{n}(r)\psi_{k}(r)dr\biggr)^2
\end{eqnarray}

\section{Conclusions}

The present research, which used the works of \cite{Nesvizh1,Nesvizh2,Nesvizh3,Nesvizh4,Nesvizh5}
as a foundation devised a simple manner in which to consider the energy of UCN weakly bound 
quantum states in the
Earth's gravitational field. Comparison of our results with the results of the above cited papers
is presented in the Table 1.

The obtained results were generalized using the case of Newton's gravitation law correction
by applying the RS theory \cite{RS} in the framework of a nonrelativistic approach. It was determined
that in the further development of measurement techniques of the UCN
in the Earth's gravitational field, we can more precisely
estimate the upper limit of the contribution of the RS correction to Newton's gravitational
law at a small distances. The contribution to the UCN energy spectrum from the RS correction to the formation
of their bound states was derived and the upper limit was estimated as $l_{RS}< 3.9\cdot 10^3 m$.
The ramifications of this are that the experimental methodology probably is not sensitive enough to check for a RS
type correction to Newton's gravitational law at small distances.

The second step involved the estimation of neutrons bound state formation in the
external gravitational field of a NS in comparison to the Earth's gravitational field neglecting GR,
magnetic and rotation effects. In this case the results obtained are presented in Table 2,
which presents some sample NS parameters and compares them with theoretical results obtained in
the Earth gravitational field.

It is obvious, that when considering the neutron bound state formation on the NS
surface it is necessary to apply a complete GR evaluation of the problem. The first
and transparent approach in this direction is a reconsideration of the SE solution for
the formation of a neutron bound state in the form:
\begin{eqnarray}
g^{ij}\biggl( \nabla_{i}\nabla_{j} -\Gamma^{k}_{ij}\nabla_{k}\biggr)\psi+
\frac{2m}{\hbar^2}\biggl( E - U_{eff}(r)\biggr)\psi =0
\end{eqnarray}
where $g^{ij}$ is a metric of the gravitational source (for example NS) and $U_{eff}(r)$ is
its effective external gravitational field. Here $i,j,k$ are spatial indexes. More mathematically
rigorous considerations would need to use the Dirac wave equation in the gravitational field of the NS.
Of course this approach would need to take into account magnetic fields and rotation effects, 
but is beyond the scope of
this work in which the goal was to present some preliminary theoretical estimations. Application of
such an approach can be very useful because of the potential of considering NS radial oscillation \cite{Gabler}
which can lead to changes in the gravitational field on the NS surface. 
Also, there may be some observable
effects related to neutron spectroscopy. 
In conclusion, we want to mention the paper \cite{Frank_Nosov},
where the UCN storage problem was considered using the previously mentioned
magnetic mirror in the presence of a gravitation field.

\begin{acknowledgments}
A.V.K. wants to express his deep gratitude to Naresh Dadhich and Ajit Kembhavi from the
Inter University Center for Astronomy and Astrophysics (IUCAA, Pune, India) for their invitation,
warm hospitality, many useful discussions and clarifying the physical aspects of the problem.
A.V.K. also wishes to thank the Bogoliubov Laboratory of Theoretical Physics  (Dubna, Russia)
for the invitation and warm hospitality, where this work was finished. In addition A.V.K. expresses
his thanks for many useful remarks (for example remarks about Fermi surface on the NS as a mirror for neutrons)
at the Low Energy Nuclear Physics Seminar at the Bogoliubov Lab. of Theoretical Physics (JINR, Dubna, Russia)
where this work was presented. R.A.S. and D.G. are grateful for a partial financial support for this
work by the St. Cloud State University (St. Cloud, Minnesota, USA) internal grant program.
\end{acknowledgments}


\end{document}